\documentstyle[12pt,epsf]{article}
\newcommand{\ska}[1]{\left( #1 \right) }

\newcommand{\pa}[2]{{\partial #1 \over \partial #2}}
\newcommand{\gsim}{\raisebox{0.2ex}{$\ > \kern -1.05em%
        \raisebox{-1.1ex}{$\sim$}\ $}}  
\begin{document}
\title{\bf 
Non-Perturbative Renormalization Group Analysis of\\
the Ohmic Quantum Dissipation}
\author{Ken-Ichi Aoki\footnote{e-mail : aoki@hep.s.kanazawa-u.ac.jp}
~and ~Atsushi Horikoshi
\footnote{Present address: Japan Science and Technology Corporation, 
and Department of Chemistry, Faculty of Science, Nara Women's University,
Nara 630-8506, Japan; e-mail : horikosi@cc.nara-wu.ac.jp
} \\
\\
Institute for Theoretical Physics, Kanazawa University, \\
Kakuma-machi Kanazawa 920-1192, Japan}
  \maketitle
\vspace{-100mm}
\begin{flushright}
KANAZAWA/02-06
\end{flushright}
\vspace{80mm}
\begin{abstract}
We analyze quantum tunneling with the Ohmic dissipation by 
the non-perturbative renormalization group method.
We calculate the localization susceptibility 
to evaluate the critical dissipation for the 
quantum-classical transition, and 
find considerably larger critical dissipation
compared to the previous semi-classical arguments.
\end{abstract}
\hspace{10mm}
PACS numbers: 11.10.Hi, 03.65.-w, 73.40.Gk 
\section{Introduction}
\hspace*{\parindent}
Caldeira and Leggett proposed a method to derive 
the quantum dissipative behavior from a microscopic theory,
which consists of the target system 
and the environment \cite{cl}.
The action of their model is written as follows, 
 \begin{eqnarray}
{S}[~q,\{x_{\alpha}\}]\!\!\!
&=&
\!\!\!\!\!\!\int d{t}~\left\{
\frac{1}{2}M{\dot q}^{2} - V_0(q)
+\hbox{$ \displaystyle\sum_{\alpha}$}
\left[\frac{1}{2}m_{\alpha}{\dot x_{\alpha}}^{2}
-\frac{1}{2}m_{\alpha}\omega_{\alpha}^{2}x_{\alpha}^{2}\right]
-q\hbox{$ \displaystyle\sum_{\alpha}$}C_{\alpha}x_{\alpha}
\right\}, \label{1}
\end{eqnarray}
where $q(t)$ is the variable of the target system 
in a potential $V_{0}(q)$, and 
$x_{\alpha}(t)$ is the harmonic oscillators representing the environment.
The target system is coupled linearly 
to each oscillator with strength $C_{\alpha}$.
If we set the parameters $m_{\alpha},~\omega_{\alpha}(>0),~C_{\alpha}$ 
in a suitable way, 
the dissipative term 
(for the Ohmic dissipation, $\dot{q}$) 
arises in the effective
{\it classical} equation of motion of $q(t)$
after elimination of the environmental variables 
with the proper boundary condition.
\par  
In this letter, we study the {\it quantum} mechanics of this system 
by the Euclidean path integral over $q(\tau)$ and $x_{\alpha}(\tau)$.
We integrate the variable $x_{\alpha}$
to define the effective action for the target system, 
 \begin{eqnarray}
Z&=&\frac{1}{\cal N}\int{\cal D}q~\hbox{$ \displaystyle\prod_{\alpha}$}
\int{\cal D}x_{\alpha}~
e^{-\frac{1}{\hbar}S_{\rm E}}=\frac{1}{{\cal N}_q}
\int{\cal D}q~e^{-\!\frac{1}{\hbar}
\int \!d\tau
~\left[\frac{1}{2}M{\dot q}^{2}+V_0(q)\right]
~-\frac{1}{\hbar}\Delta S[q]},\label{2}
 \end{eqnarray}
where $S_{\rm E}$ is the Euclidean action
 \begin{eqnarray}
{S_{\rm E}}[~q,\{x_{\alpha}\}]\!\!\!
&=&
\!\!\!\!\!\!\int d{\tau}~\left\{
\frac{1}{2}M{\dot q}^{2} + V_0(q)
+\hbox{$ \displaystyle\sum_{\alpha}$}
\left[\frac{1}{2}m_{\alpha}{\dot x_{\alpha}}^{2}
+\frac{1}{2}m_{\alpha}\omega_{\alpha}^{2}x_{\alpha}^{2}\right]
+q\hbox{$ \displaystyle\sum_{\alpha}$}C_{\alpha}x_{\alpha}
\right\},
\label{3}
 \end{eqnarray}
$\cal N$ and ${\cal N}_q$ are normalization constants and 
$\Delta S[q]$ is a term generated by the quantum effects 
of environment $x_{\alpha}$.
Caldeira and Leggett studied the influence of $\Delta S$ on 
the quantum tunneling of the target system $q$ 
and they found that the quantum tunneling is suppressed by
the Ohmic dissipation \cite{cl}.
Following their analysis, many theoretical works have been done 
and the localization (quantum-classical) transition 
has been suggested to occur
\cite{cha,bm,fz,fiss,weiss,lkppk,zlp}.
The validity of their results have been suggested 
by elaborate experiments \cite{fvdz}.
However, the theoretical works so far 
(instanton, perturbation, etc.) depend on 
some expansions with respect to small parameters 
which are not always valid. 
For example,  
for evaluation of the 
tunneling rate through the double well barrier, 
 \begin{eqnarray}
V_0(q)=-\frac{1}{2}M\omega_0^2 q^2+\lambda _0 q^4,\label{4}
 \end{eqnarray}
the dilute gas instanton calculation is 
valid only in $\lambda _0\to 0$ region \cite{ahtt}.
\par
In this letter we adopt 
the non-perturbative renormalization group (NPRG) 
method \cite{wk,wh,ap}
to analyze the dissipative quantum tunneling 
in the Caldeira-Leggett model.
Since the NPRG is formulated without any series expansion,
it is a powerful tool to analyze the 
non-perturbative features
of quantum field theory,
such as the phase structure of the theory \cite{aoki}.
For the quantum tunneling problem, 
the NPRG method can reproduce the exact 
tunneling rate through the double well barrier (Eq. (\ref{4}))
in the wide parameter $\lambda _0$ region \cite{ahtt,kt,za}.
Therefore the NPRG is expected to be 
effective in the non-perturbative analysis 
of the dissipative quantum tunneling.
We can discuss the possibility of 
the localization (quantum-classical) transition
by investigating the ``phase structure'' 
of the dissipative quantum mechanics.
The NPRG also helps us to interpret 
the effects of $\Delta S$
as effective infrared or ultraviolet cutoffs 
for the quantum fluctuation, 
and we can readily understand the dissipation effects 
as suppression or enhancement of the
quantum features of the effective target system \cite{ah}. 
\par
\section{Dissipative effective action}
\hspace*{\parindent}
The effective interaction $\Delta S$ 
defined in Eq. (\ref{2}) takes the following form,
\begin{eqnarray}
\Delta S[q]&=&-~\int d\tau\!\int ds~q(\tau)
~\alpha(\tau -s)~q(s). \label{5}
\end{eqnarray}
The non-local coupling coefficient is given by
\begin{eqnarray}
\alpha(\tau -s)&=&
\int \frac{d\omega}{2\pi}
~\sum_{\alpha}~\frac{C^2_{\alpha}}{2 m_{\alpha}}
\frac{1}{\omega^2+\omega^2_{\alpha}}~e^{i\omega(\tau -s)}\nonumber\\
&=&\!\int^{\infty}_{0} 
\frac{d\omega}{2\pi}~J(\omega)~e^{-\omega|\tau -s|},\label{6}
\end{eqnarray}
where we have introduced the spectral density function $J(\omega)$ 
characterizing the environment,
\begin{eqnarray}
J(\omega)&=&\sum_{\alpha}
~\frac{C^2_{\alpha}}{4 m_{\alpha}\omega_{\alpha}}
~(2\pi)~\delta(\omega-\omega_{\alpha}).\label{7}
\end{eqnarray}
If we set $J(\omega)=\eta~\!\omega$, the Ohmic dissipation term
$-\eta\dot{q}$ arises 
in the effective classical equation of motion of $q(t)$
with the proper boundary condition 
of the environmental variables \cite{fiss,lkppk}.
\par
Generally $\Delta S$ consists of 
the local component $\Delta S_{\rm L}$ and 
the non-local component $\Delta S_{\rm NL}$.
We identify the component $\Delta S_{\rm NL}$ 
as the dissipation term,
since it corresponds to the odd power term in the Fourier transform
and then it is related to the time reversal symmetry breaking.
On the other hand,
the component $\Delta S_{\rm L}$ consists of 
the even power term in the Fourier transform 
and then it does not contribute the time reversal symmetry
breaking in the effective classical equation of motion, 
that is, it is irrelevant to the dissipation.
Therefore, we introduce a suitable conterterm 
in the Euclidean action $S_{\rm E}$ 
to cancel the component $\Delta S_{\rm L}$ \cite{cl}.
\par
For the Ohmic dissipation $J(\omega)=\eta~\!\omega$,
we identify the dissipative part as follows:
\begin{eqnarray}
\Delta S_{\rm NL}[q]&=&\!\frac{\eta}{4\pi}\int d\tau\!\int ds~
\frac{\left(~q(\tau)-q(s)~\right)^2}{|\tau -s|^2}.\label{8}
 \end{eqnarray}
It should be noted that 
the dissipation term, $\Delta S_{\rm NL}[q]$,
never breaks the time reversal symmetry in the action level.
The actual dissipative effect ($-\eta\dot{q}$) arises 
only in the classical equation of motion level \cite{fiss,lkppk}.
Using the Fourier transform of $q(\tau)$,
$\Delta S_{\rm NL}$ is also represented as \cite{lkppk}
\begin{eqnarray}
\Delta S_{\rm NL}[q]
&=&\!\frac{1}{2}~\!\!\int \frac{d\omega}{2\pi}
~\eta~|\omega|~
\tilde{q}~\!(\omega)~\!\tilde{q}~\!(-\omega).\label{9}
 \end{eqnarray}
Note that $\eta$ is a dimensionful parameter, 
[$\eta$]=[M$\omega$]=[mass]$\times$[time]$^{-1}$. 
We study the influence of the dissipation term
$\Delta S_{\rm NL}$ on 
the quantum behaviors of $q(\tau)$. 
\section{NPRG equation with the Ohmic dissipation}
\hspace*{\parindent}
Now let us proceed to the NPRG analysis of 
the Caldeira-Leggett model \cite{comment1}.
Originally the NPRG method has been formulated and used mainly 
in the statistical mechanics or the quantum field theory 
to analyze the critical phenomena \cite{wk,wh,ap,aoki}.
Recently, the NPRG method has been found even effective in the
quantum mechanics particularly for the 
non-perturbative analysis \cite{ahtt,kt,za}.
\par
We derive the NPRG equation for the quantum mechanics 
with the dissipation term.
In the NPRG method, the theory is defined 
as an effective theory with 
a high frequency cutoff $\Lambda$, 
and is described by 
the Wilsonian effective action 
$S_{\Lambda}[q]$,
which changes with respect to the 
scale $\Lambda$.
We now employ the local potential approximation (LPA),
\begin{eqnarray}
S_{\Lambda}[q]&=&\int d\tau \left[
~\frac{1}{2}M\dot{q}^2 +
\frac{\eta}{4\pi}\int ds
\frac{\left(~q(\tau)-q(s)~\right)^2}{|\tau -s|^2}
+ V_{\Lambda}(q)
~\right],\label{10}
 \end{eqnarray}
where the effective action is limited to have only local potential 
term, the Wilsonian effective potential $V_{\Lambda}(q)$, 
in addition to the fixed kinetic term
and the fixed dissipation term.
Both the particle mass $M$ and the dissipation strength $\eta$ 
are constant and 
only the Wilsonian effective potential $V_{\Lambda}(q)$
is changed as the scale $\Lambda$ is lowered.
Note that the LPA is the leading order of the 
derivative expansion of $S_{\Lambda}[q]$ 
and has nothing to do with 
the prescription that we ignore the local part 
$\Delta S_{\rm L}$ in the component $\Delta S$.  
The LPA means that any quantum correction to
the derivative coupling is ignored,
but does not mean that any quantum correction from
the derivative coupling is ignored.
We carry out the path integration over the highest frequency
degrees of freedom of $\tilde{q}(\omega)$, 
$\Lambda-\Delta \Lambda<|\omega|\le\Lambda$,
up to the one-loop order, 
 \begin{eqnarray}
V_{\Lambda-{\mit\Delta}\Lambda}(q)&=&V_{\Lambda}(q)
+\hbar~{\int_{\rm \Lambda-\Delta \Lambda}^{\Lambda}}
\frac{d\omega}{2\pi}~{\rm log}\left(
M\omega^2+\eta~\!|\omega|
+\frac{\partial^2 V_{\Lambda}}{\partial q^2}
\right), \label{11}
 \end{eqnarray}
where we have omitted a proper constant term which should make the 
argument of the logarithm to be dimensionless, since it contributes 
only to the $\it q$-independent part of the effective action.
Taking the limit ${\mit\Delta}\Lambda\to 0$ in Eq. (\ref{11}),
we have the NPRG equation
\begin{eqnarray}
 \Lambda\pa{V_{\Lambda}}{\Lambda}&=&
-~\!\frac{\hbar}{2\pi}~\!\Lambda
  ~\!\log\ska{1+\frac{1}{\Lambda}\frac{\eta}{M}
+
\frac{1}{\Lambda ^2}\frac{1}{M}\frac{\partial ^2
  V_{\Lambda}}{\partial q ^2} },\label{12}
 \end{eqnarray}
which should be called the LPA Wegner-Houghton equation with dissipation. 
This is because in the $\eta\to0$ limit Eq. (\ref{12}) becomes 
the ordinary LPA Wegner-Houghton equation \cite{wh,ap}, which 
describes the change of the Wilsonian effective potential $V_{\Lambda}(q)$
with respect to the scale $\Lambda$.
The extra term $\eta /(\Lambda M)$
originates from the dissipation term $\Delta S_{\rm NL}$.
The NPRG equation (Eq. (\ref{12})) is obtained with 
the approximation LPA; the leading order of the 
derivative expansion of $S_{\Lambda}[q]$. 
The derivative expansion does not rely on
any small parameter which controls the series expansion,
and therefore the effectiveness of the NPRG equation 
is expected to be free from the smallness of the parameters
in the system.  
\section{Analysis of Dissipative Quantum Tunneling}
\hspace*{\parindent}
We regard the system variable $q$ 
as a {\it macroscopic} collective coordinate and 
discuss how the quantum mechanical behavior of the 
macroscopic  coordinate $q$ 
is affected by the environmental effects.
As a typical phenomenon, the quantum tunneling of $q$
has been analyzed by many authors \cite{cl,cha,bm,fiss}.
We set the bare potential of $q$ as 
the double well type (Eq. (\ref{4})).
In this system with vanishing $\eta$, the coordinate $q$ tunnels 
through the potential barrier and it oscillates 
between two wells.
The main question is whether it also oscillates or not even in the
$\eta\ne 0$ quantum dissipative dynamics. 
Such an oscillating behavior is particularly called 
the {\it macroscopic quantum coherence} and 
the inclination of ceasing the oscillation  
corresponds to {\it decoherence} \cite{weiss}.  
\par
Now we briefly summarize the previous results    
obtained for the macroscopic quantum coherence in double well
potential systems with the Ohmic dissipation.
Usually the first energy gap $\Delta E=E_1-E_0$ is calculated 
as the physical quantity because it corresponds to the 
tunneling amplitude between two wells.
Caldeira-Leggett evaluated it
by using the semi-classical (instanton)
approximation which is valid for $\lambda _0\to 0$ 
and the perturbation with respect to $\eta$, 
and then they found the dissipation term   
$\Delta S_{\rm NL}$ suppresses the quantum tunneling \cite{cl}.
Renormalization group analyses have been done for the 
instanton gas system within  
the dilute gas approximation which 
is valid also for 
$\lambda _0\to 0$ and $\eta\to 0$ region \cite{cha,bm}.
They predict a remarkable phenomenon that     
$\Delta E$ vanishes
at a critical value $\eta_c=2\pi\hbar\lambda_0/(M\omega_0^2)$, 
where the  
decoherence occurs.
This phenomenon is often called 
the {\it quantum-classical transition} 
because the system looks classical in a limited sense 
that the quantum tunneling does not occur \cite{comment2}.  
This is nothing but the spontaneous $Z_2$ symmetry breaking. 
In these works the energy gap $\Delta E$
was treated as the order parameter of the transition. 
\par
These results are obtained by using the semi-classical 
approximation and/or the perturbation theory with respect to $\eta$.
Their reliability depend on the 
smallness of the couplings $\lambda_0$ and $\eta$. 
Therefore, to get more general and reliable results 
free of such limitation,
we must employ an analyzing tool which does not 
need the series expansion with respect
to any couplings.
We expect our method using 
the LPA Wegner-Houghton equation with dissipation (Eq. (\ref{12})) 
will work best for the large coupling region,
because the employed approximation, LPA, is the leading order 
of the derivative expansion and does not depend 
any small parameter. 
In fact, it has been found that 
the NPRG analysis of quantum tunneling 
based on the LPA Wegner-Houghton equation 
works very well in the wide parameter 
($\lambda_{0}$) region \cite{ahtt}.
\par
We analyze the dissipative quantum tunneling 
by solving the NPRG equation (Eq. (\ref{12})) numerically.
This equation is a two-dimensional partial differential equation 
for the Wilsonian effective potential $V_{\Lambda}(q)$ 
with respect to $\Lambda$ and $q$. 
The initial condition of the potential
is the bare potential $V_{\Lambda_{0}}(q)=
-\frac{1}{2}M\omega_0^2 q^2+\lambda _0 q^4$ 
at the initial cutoff $\Lambda_0$.
We solve the differential equation toward the infrared limit 
$\Lambda\to 0$ and finally obtain the physical effective potential 
$V_{\rm eff}(q)=\frac{1}{2}M\omega_{\rm eff}^2 q^2
+\lambda _{\rm eff} q^4+\cdot\cdot\cdot$.
We can exploit the physical information of the quantum system 
from the effective potential $V_{\rm eff}(q)$.
First of all, the frequency squared starting with the negative value  
$\omega_{\rm \Lambda_0}^{2}=-\omega_0^{2}$ 
finally reaches a positive value (for small $\eta$), 
$\omega_{\rm eff}^{2}>0$, that is, the effective potential $V_{\rm eff}(q)$ 
becomes a single well form.
It means that due to the quantum tunneling, the $Z_2$ symmetry
(parity) does not break spontaneously, and the expectation value of
$q$ is vanishing.
The tunneling effects are automatically incorporated 
when we integrate (solve) the NPRG equation toward the infrared.
\par
It should be noted that 
for non-vanishing $\eta$, 
the simple equivalence between 
the energy gap $\Delta E$ and the effective frequency 
$\omega_{\rm eff}$, $\Delta E=\hbar\omega_{\rm eff}$ \cite{ahtt}, 
does not hold.   
The long range ($\tau\to\infty$) behavior of the two point function 
 \begin{eqnarray}
\lim_{\tau\to\infty}\left\langle \Omega \right|
{T}\hat{q}(\tau)\hat{q}(0)
\left| \Omega \right\rangle
=
\lim_{\tau\to\infty}
\int \frac{d\omega}{2\pi}e^{i\omega\tau}
\frac{\hbar}{M\omega^2+\eta|\omega|+M\omega^2_{\rm eff}}, \label{13}
 \end{eqnarray}
is actually independent of $\omega_{\rm eff}$ and is determined by 
$\eta$ with power damp behavior even in case of infinitesimal $\eta$.
This singular behavior comes from the nature of the environmental 
degrees of freedom where the harmonic oscillator frequencies are
assumed to distribute continuously up to zero ($\omega_{\alpha}=0$). 
Even if there is an infrared cutoff, it does not change the situation 
much and the lowest frequency environment dominates the long range 
correlator of the target system. 
In this sense the semi-classical arguments 
calculating the first energy gap $\Delta E=E_1-E_0$
as the order parameter of the quantum-classical transition
in the quantum dissipative systems 
seems somewhat doubtful,
because the correspondence between
the tunneling amplitude and 
the first energy gap $\Delta E$ comes form 
the long range behavior of the transition amplitude.
\par
Here, instead of analyzing the long range correlator, we adopt 
another physical quantity to describe the possible quantum-classical 
transition, the localization susceptibility. 
We add a source term 
 \begin{eqnarray}
J\tilde{q}_{T}(0)\equiv
J\int^{T/2}_{-T/2}d\tau ~q(\tau)  \label{14}
 \end{eqnarray}
to the Euclidean action $S_{\rm E}$ (Eq. (\ref{3})).
Then the localization susceptibility $\chi$ is defined by
 \begin{eqnarray}
\chi\equiv
\left.\frac{d \langle q(\tau) \rangle_{J}}{d J}\right|_{J=0}=
\lim_{T\to \infty}\frac{1}{T}
\left.\frac{d \langle \tilde{q}_{T}(0) \rangle_{J}}{d J}\right|_{J=0}
=\lim_{T\to \infty}\frac{1}{T}
\left.\frac{d^2 \log Z}{d J^2}\right|_{J=0}, \label{15}
 \end{eqnarray}
and it is exactly given by the effective potential as follows:
 \begin{eqnarray}
\chi=\left(\left.\frac{d^2 V_{\rm eff}(q)}{d q^2}\right|_{q=q_{\rm min}}
\right)^{-1}=
\frac{1}{M \omega^2_{\rm eff}}, \label{16}
 \end{eqnarray}
where $q_{\rm min}$ is a minimum of $V_{\rm eff}(q)$ and 
it is actually vanishing in the sub-critical region. 
These definitions are valid also for non-vanishing $\eta$.
If we find a divergent and scaling behavior of $\chi$
toward $\eta=\eta_c$, we may conclude that 
the localization transition occurs at $\eta_c$ and it is the second
order phase transition.
\par
\begin{figure}[htb]       %
\hspace{0mm}
 \parbox{78mm}{
 \epsfxsize=78mm      %
 \epsfysize=78mm
  \leavevmode
\epsfbox{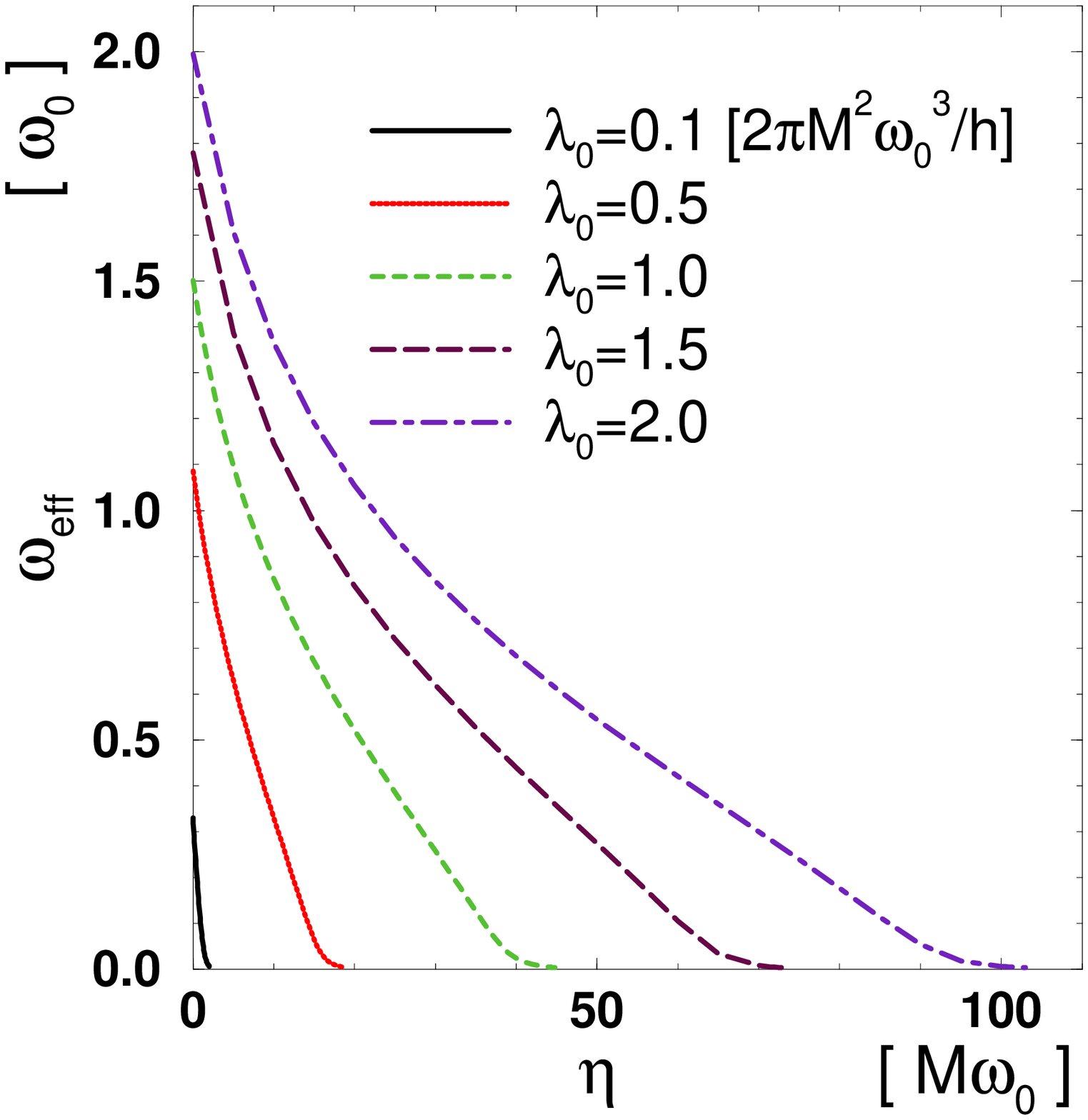}
\vspace{-5mm}
\caption{ The effective frequency $\omega_{\rm eff}$
with initial cutoff $\Lambda_0=10^4[\hbar\omega_0]$. }
 \label{fig:ohm} 
}
\hspace{0mm} 
\parbox{78mm}{
 \epsfxsize=78mm      %
 \epsfysize=78mm
 \leavevmode
\epsfbox{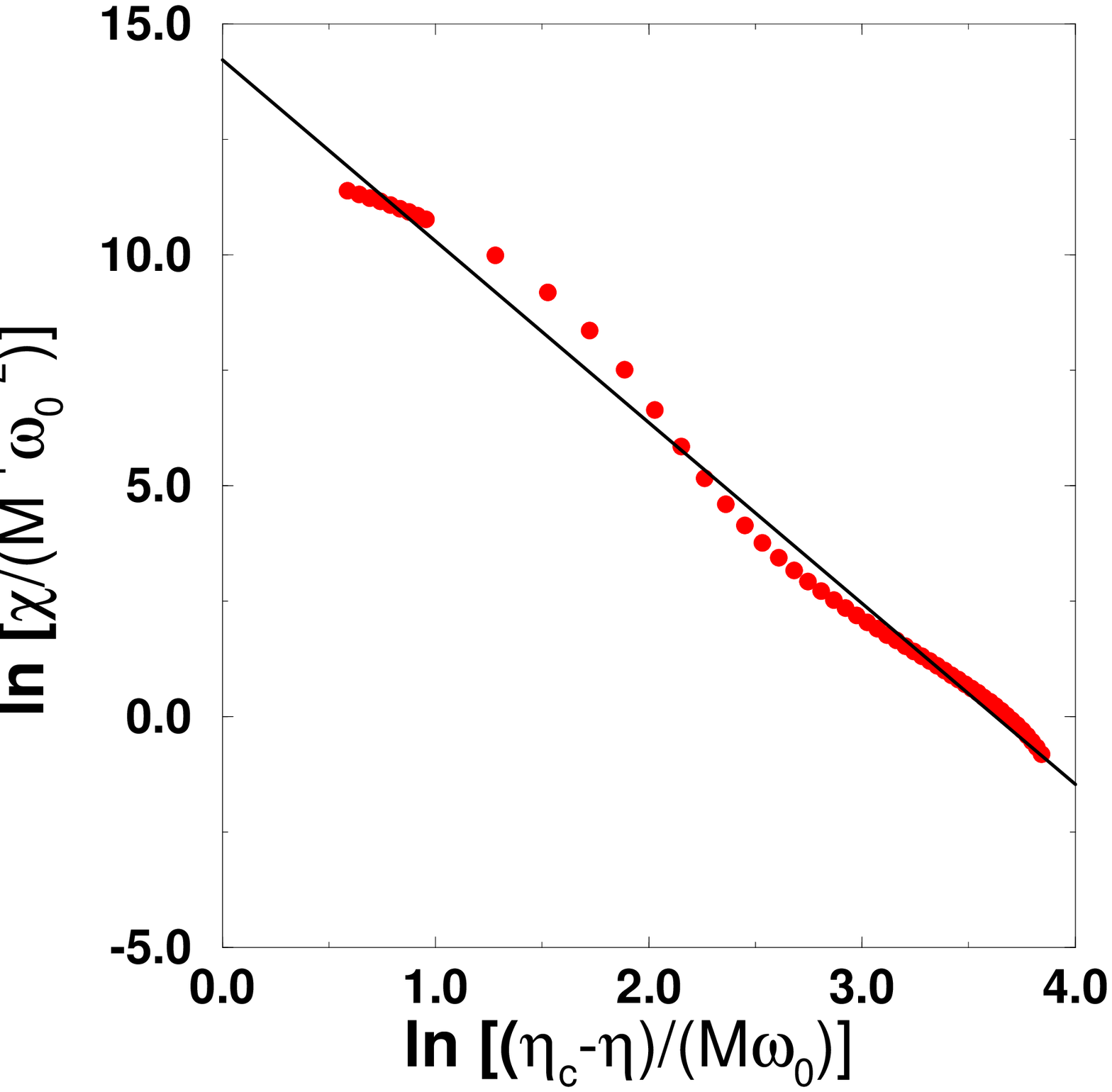}
\vspace{-5mm}
\caption{Critical scaling fit of 
localization susceptibility $\chi$ for 
$\lambda_0=1.0[M^2\omega_0^3/\hbar]$. }
\label{fig:sus}
 }
\end{figure}
\par
The results 
are shown in Fig.\ref{fig:ohm} for several values of $\lambda_0$.
In previous works for $\eta=0$ system \cite{ahtt,kt,za},
the NPRG analyses have been found to work excellently 
in these $\lambda _0$ region ($\lambda_0\gsim 0.1[M^2\omega_0^3/\hbar]$)
while the dilute gas instanton approximation 
does not work at all there.
Therefore, we expect the NPRG analysis also works well 
for these $\lambda _0$ values
even in $\eta\ne 0$ systems.
We find that for every value of $\lambda_0$, $\omega_{\rm eff}$ decreases as 
$\eta$ becomes large.
This $\eta$ and $\lambda_0$ dependence of $\omega_{\rm eff}$ is 
naturally understandable by considering 
the dissipation effect in Eq. (\ref{12}).
The quantum fluctuation of the dynamical variable 
$q$ is suppressed bellow
the effective infrared cutoff $\omega_{\rm IR}=\eta/M$. 
Therefore, the larger $\eta$ causes 
the stronger cutoff effect and results in
the smaller $\omega_{\rm eff}$.
For larger $\lambda_0$, the effective frequency 
of the system is larger, and therefore 
larger $\eta$ is required to effectively cutoff the quantum
fluctuation.
These behaviors are qualitatively consistent
with those of $\Delta E$ obtained by 
the instanton approximation.
\par
\begin{figure}[htb]
\begin{center}
\epsfysize=78mm
\epsfxsize=78mm
\leavevmode
\epsfbox{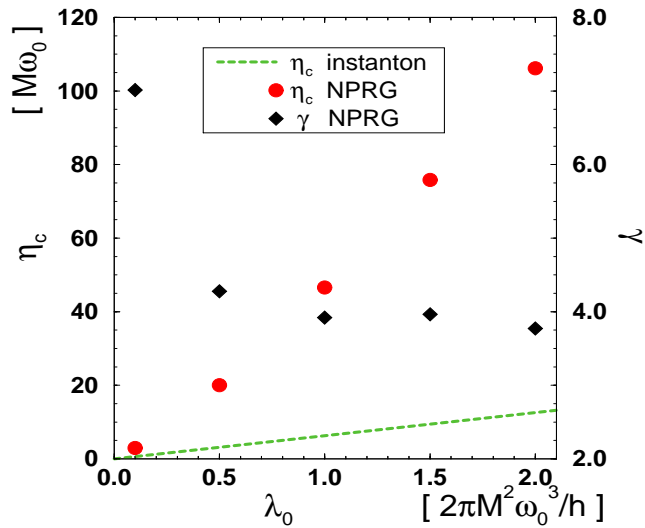}
\vspace{-5mm}
\caption{$\lambda_0$ dependence of
critical dissipation $\eta_c$ and critical exponent $\gamma$.}
 \label{fig:eg}
\end{center}
\end{figure}

Then, what happens for large $\eta$ case?
The expected phenomenon is the quantum-classical transition
characterized by complete disappearance of the tunneling.
The NPRG method may not work precisely in the 
$\omega_{\rm eff}/\omega_{0}\to 0$ region because Eq. (\ref{12})
becomes singular there and the numerical error increases.
Therefore, according to the standard technique of 
analyzing the critical phenomena,
we evaluate the critical dissipation $\eta_c$ 
from the diverging behavior of the 
localization susceptibility.
We fit $\chi(\eta)$ with the critical exponent form
 \begin{eqnarray}
\chi=C |\eta-\eta_c|^{-\gamma}.\label{17}
 \end{eqnarray}
We show an example of fit for $\lambda_0=1.0[M^2\omega_0^3/\hbar]$ 
in Fig.\ref{fig:sus}.
We conclude that the localization susceptibility exhibits a divergent 
behavior with a power scaling, which indicates the second order phase 
transition of localization. 
We obtain the critical dissipation $\eta_c$ and 
the critical exponent $\gamma$ for these $\lambda_0$, which 
are plotted in Fig.\ref{fig:eg}. 
The previous analysis using the dilute gas instanton gives a simple 
relation $\eta_c=2\pi\hbar\lambda_0/(M\omega_0^2)$ 
which is also shown in Fig.\ref{fig:eg} 
for comparison. 
The NPRG results for $\eta_c$ are systematically larger compared to 
those of the instanton.
As for the critical exponent $\gamma$, we observe the universality 
property except for the smallest $\lambda_0=0.1[M^2\omega_0^3/\hbar]$, 
where NPRG results 
may not be reliable.
\par
\par
The NPRG works very well 
in the large $\lambda_0$ region \cite{ahtt,kt,za}. 
Strictly speaking, tunneling phenomena are no longer characteristic 
in such 
parameter region. 
It can be easily seen by evaluating a dimensionless quantity 
$r=8\sqrt{2}\hbar\lambda_0/(M^2\omega_0^3)$, 
which is the ratio of the zero-point energy of the particle 
at each well with $\eta=0$
to the height of the inter-well barrier. 
Tunneling phenomena are dominant for $r\ll 1$ case. 
For $\lambda_0=0.1[M^2\omega_0^3/\hbar]$, $r\sim 1.1$ is obtained.  
Therefore, our analysis in the large $\lambda_0$ region 
($\lambda_0\gsim 0.1[M^2\omega_0^3/\hbar]$)
rather claims suppression of the quantum oscillation over 
the small inter-well barrier.
The critical phenomenon that we observed here is 
induced by a destruction of the quantum coherence
and can be interpreted as spontaneous $Z_2$ symmetry 
breaking-like phenomenon in quantum mechanics.
\section{Summary}
\hspace*{\parindent}
The Caldeira-Leggett model 
for dissipative quantum mechanics
was analyzed by means of the NPRG method.
We derived the LPA Wegner-Houghton equation 
with dissipation, which is 
the local potential approximated 
NPRG equation with the Ohmic dissipation term.
We applied it to analysis of 
dissipative quantum tunneling (or dissipative quantum oscillation)
and investigated whether the quantum-classical transition
occurs or not.
Since the first energy gap $\Delta E=E_1-E_0$ is
unsuitable as the order parameter of the dissipative 
phase transition, 
we calculated the localization susceptibility $\chi$
from the effective potential $V_{\rm eff}(q)$.
The observed $\chi$ diverges
toward $\eta=\eta_c(\lambda_0)$ 
with the critical scaling, 
and therefore we concluded that 
$\eta_c(\lambda_0)$ is the critical dissipation 
for the quantum-classical transition and
this transition seems the second order phase transition.
The values of the critical dissipation $\eta_c(\lambda_0)$
are rather larger than the values
obtained by the semi-classical analyses, where
the first energy gap $\Delta E$ is used as 
the order parameter of the quantum-classical transition.
As for the non-Ohmic dissipations, the analogous systematic study using 
the NPRG equation will be reported elsewhere \cite{ah}.
There must remain, however, many fundamental and subtle problems 
to be studied about the notion of quantum-classical transition
and the critical dissipation.
The NPRG method will be effective also for those problems.
\section*{Acknowledgement}
 \pagestyle{myheadings}                      
\hspace*{\parindent}
We would like to thank I. Sawada for fruitful and encouraging 
discussions and suggestions. 
K.-I. Aoki is partially supported by the Grant-in Aid for
Scientific Research (\#12874029) from the
Ministry of Education, Science and Culture.


\end{document}